\documentclass[aps,pre,superscriptaddress,twocolumn,longbibliography]{revtex4}

\usepackage{graphicx,epstopdf,url}
\usepackage[colorlinks=true,urlcolor=blue,citecolor=blue,linkcolor=blue,urlcolor=blue]{hyperref}

\usepackage[active]{srcltx}

\begin{document}

\title{
Multistability of carbon nanotube packings on flat substrate
}

\author{A. V. Savin}
\email{asavin@center.chph.ras.ru}
\affiliation{
N.N. Semenov Federal Research Center for Chemical Physics,
Russian Academy of Sciences (FRCCP RAS), Moscow, 119991, Russia\\
}
\affiliation{
Plekhanov Russian University of Economics, Moscow, 117997 Russia
}

\begin{abstract}
It is shown by the method of molecular dynamics using a chain model that a multilayer packaging
of identical single-walled carbon nanotubes with a diameter of $D>2.5$~nm located on a flat substrate
is a multistable system. The system has many stationary states, which are characterized by the
portion of collapsed nanotubes. The thickness of the package monotonically decreases with
an increase in the portion of such nanotubes. For nanotubes with a chirality index (60,0),
depending on the portion of collapsed nanotubes, the thickness of the 11-layer package can
vary from 12 to 36~nm. All stationary states of the package are stable to thermal fluctuations at $T=300$~K.
The transverse compression of the package is not elastic; due to the collapse of a part
of the nanotubes, it only transfers the package from one stationary state to another
with a smaller thickness.
 \\ \\
\end{abstract}


\maketitle

\section{Introduction}
Carbon nanotubes (CNTs) are macromolecules with a cylindrical form with a diameter starting
from 0.4~nm and lengths up to few microns.
Similar structures were detected for the first time during the thermal decomposition
of carbon monoxide on an iron contact
\cite{Radushkevich52}.
CNTs as such were obtained considerably later as side products of C$_{60}$ fullerene synthesis
\cite{Iijima91}.
Currently, CNTs attract interest due to their unique properties
\cite{qian2002}.
CNTs with desired geometric properties (i.e., with the required diameter, length, and chirality)
can be readily synthesized
\cite{di2016,bai2018}
and used to prepare bundles of parallel CNTs
\cite{liu2003,li2005}.
Such materials, also referred to as CNT forests or arrays,
feature even more superior mechanical properties, compared to isolated CNTs,
due to van der Waals interactions between them
\cite{rakov2013}.

A number of computational approaches have been developed for the study of
structures based on CNTs in addition to the well-known molecular dynamics method. Deformation
mechanisms of CNT forest have been studied using mesoscopic modeling in
\cite{Wittmaack2018,Wittmaack2019}.
A continuum thin shell theory is capable of describing
large deformations of CNTs
\cite{Yakobson1996}.
Mechanical properties of CNTs
under transverse loading have been studied in \cite{Saether2003}.
In the review
\cite{Raffi2016}
the power of the nonlocal beam, plate, and shell theories
in modeling mechanical properties of nanomaterials is described. The application
of the continuum beam theory has been demonstrated in
\cite{Harik2001}.
Simulation of the mechanical properties and failure of the CNT bundles have been performed
using a nonlinear coarse-grained stretching and bending potentials \cite{Ji2019}.

CNTs of a diameter greater than a threshold value can exist in circular and in collapsed forms
due to the competition between CNT wall bending and van der Waals interactions
\cite{chopra1995,Chang2008,Impellizzeri2019,cui2019,Maslov2020}.
Unlike dense materials, CNT crystal can demonstrate very high compressibility in the
elastic region. Practically, CNT bundles can be used, e.g., for protection against shocks
and vibrations \cite{Cao2005,Rysaeva2020}.

In this work, we consider CNT arrays on a flat substrate formed by the surface
of hexagonal boron nitride (h-BN) crystal. Using chain model offered in
\cite{savin2015prb}
and modified for the CNT bundle in
\cite{Korznikova2019},
we will show that the multilayer
packaging of single-walled nanotubes on a flat substrate is a multistable system.
The stationary states of the packages differ slightly in energy, but they can differ several
times in thickness.

\section{The model}
An array (a bundle) of CNTs can be conveniently described with a 2D model of a system
of cyclic molecular chains
\cite{savin2015prb,savin2017cms}.
For a single-walled
CNT with a zigzag structure (with chirality index $(m,0)$), the chain model describes the nanotube's
transversal cross section that forms a ring-shaped chain of $N=2m$ effective particles
corresponding to longitudinal lines of atoms in the nanotube.

Under plane strain conditions, the cross section of a CNT completely determines its deformed
state. Then a bundle of defect-free CNTs can be represented by a set of their cross sections,
which significantly reduces the number of considered degrees of freedom.
This model was previously successfully used in modeling the scrolls of graphene nanoribbons
\cite{savin2015prb},
windings of nanoribbons around CNTs
\cite{savin2017cms},
analysis of mechanical
properties of CNT bundle upon uniaxial and biaxial lateral compression
\cite{Korznikova2019}.
\begin{figure}[tb]
\begin{center}
\includegraphics[angle=0, width=1.0\linewidth]{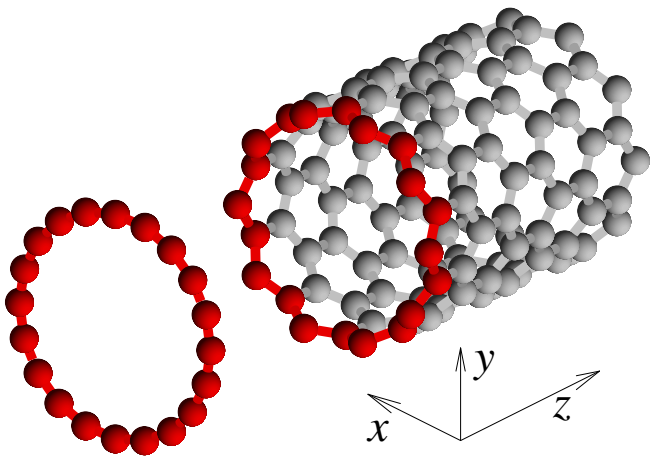}
\end{center}
\caption{\label{fig01}\protect
A scheme for constructing a 2D chain model of single-walled zigzag
carbon nanotube.
Nanotube with chirality index of $(m,0)$ with $m=10$
and corresponding cross section of $N=2m$ carbon atom is shown.
}
\end{figure}

For definiteness, a bundle of straight, single-walled zigzag CNTs with chirality $(m,0)$
oriented along the $z$-axis of the Cartesian coordinate system, as shown in Fig.~\ref{fig01},
is considered. The cross sections of such CNT contain $N=2m$ carbon atoms,
each of which represents a straight chain of atoms normal to the $(x,y)$-plane. Each
atom in the cross section of a CNT has two degrees of
freedom, namely, coordinates on the $(x,y)$-plane.

The Hamiltonian for single CNT cross section, with the atoms numbered by the index $n=1,...,N$,
can be written in the form
\begin{eqnarray}
H=\sum_{n=1}^{N}[\frac12M(\dot{\bf u}_n,\dot{\bf u}_n)+V(R_n)+U(\theta_n)
\nonumber\\
+W_0(y_n)+\frac12\sum^N_{l=1\above 0pt |l-n|>4} W_1(r_{nk})],
\label{f1}
\end{eqnarray}
where the 2D vector ${\bf u}_n=(x_n,y_n)$ defines the coordinates of the $n$-th particle
in the cycle chain,
and $M=12m_p$ is the carbon atom mass ($m_p=1.6603\cdot 10^{-27}$~kg is the proton mass).

The potential
\begin{equation}
V(R)=\frac12K(R-R_0)^2,
\label{f2}
\end{equation}
describes the longitudinal chain stiffness, where $K$ is the interaction stiffness,
$R_0$ is the equilibrium bond length (chain period), and
$R_n=|{\bf v}_{n}|$ is the distance between adjacent $n$ and $n+1$ nodes
(vector ${\bf v}_n={\bf u}_{n+1}-{\bf u}_n$).

The potential
\begin{equation}
U(\theta)=\epsilon_\theta[1+\cos(\theta)],
\label{f3}
\end{equation}
describes the bending stiffness of the chain, where $\theta$ is the angle between two adjacent bonds,
cosine of the $n$-th valence angle $\cos(\theta_n)=-({\bf v}_{n-1},{\bf v}_n)/R_{n-1}R_n$.

The parameters of potentials (\ref{f2}) and (\ref{f3}) were determined in
\cite{savin2015prb}
by analyzing dispersion curves for a graphene nanoribbon:
longitudinal stiffness $K=405$~N/m, chain period $R_0=r_c\sqrt{3}/2$
(where $r_c=1.418$~\AA~ is the length of C--C valence bond in the graphene sheet,
$R_0=1.228$~\AA),
and energy $\epsilon_\theta=3.5$~eV. The diameter of an isolated $(m,0)$ nanotube
is $D=R_0/\sin(\pi/2m)\approx 2m R_0/\pi$.

Potential $W(r_{n,k})$ describes weak non-covalent interactions between remote nodes $n$ and $k$
of the chain, where $r_{n,k}=|{\bf u}_k-{\bf u}_n|$ is the distance between the nodes.
This potential was also used to describe the interaction between nodes of different chains
(different nanotubes). The potential of non-covalent interaction between chain nodes can be described
with high accuracy
\cite{savin2019prb}
by the (5,11) Lennard-Jones potential
\begin{equation}
W_1(r)=\epsilon_1[5(r_0/r)^{11}-11(r_0/r)^5]/6, \label{f4}
\end{equation}
with equilibrium bond length $r_0=3.607$~\AA~ and interaction energy $\epsilon_1=0.00832$~eV.

In the chain Hamiltonian (\ref{f1}), potential $W_0(y)$ describes the interaction between
a chain node and the substrate formed by the flat surface of a molecular crystal.
In modeling, we take that the substrate surface coincides with the plane $y=0$.
To determine this potential, we established numerically the dependence of the energy
of interaction between a carbon atom and the substrate on its distance to the surface plane $y$.
Calculations
\cite{savin2019prb,savin2019pss}
showed that the energy $W_0(y)$
of interaction with the substrate can be described with good accuracy by the $(k,l)$
Lennard-Jones potential:
\begin{equation}
W_0(h)=\varepsilon_0[k(h_0/h)^l-l(h_0/h)^k]/(l-k), \label{f5}
\end{equation}
where $l>k$. Potential (\ref{f5}) has the minimum $W_0(h_0)=-\epsilon_0$, where $\epsilon_0$
is the energy of bonding between a carbon atom and a substrate, and $h_0$ is the equilibrium
distance from the surface plane of the substrate. 
{
Potential (\ref{f5}) allows us to describe the interaction of a carbon atom with 
flat surfaces of molecular crystals of ice  I$_h$, graphite, silicon carbide $6H$-SiC, 
silicon and silver \cite{savin2019pss}.
We will use the flat surface of hexagonal boron nitride (h-BN) as a substrate,
since it is the ideal substrate for graphene nanoribbon and nanotube \cite{wang2017,ling2019}.
For the h-BN surface interaction energy is $\epsilon_0=0.0903$~eV, the equilibrium distance is
$h_0=3.46$~\AA, and the exponents are $l=10$ and $k=3.75$.

The potentials (\ref{f4}) and (\ref{f5}) are obtained as the sums of the Lennard-Jones potentials 
(6,12) describing the Van der Waals interactions of pairs of atoms. 
Therefore, in 2D model they describe Van der Waals interactions.
The bending stiffness of nanotubes is described by the potentials (\ref{f2}) and (\ref{f3}).
Due to the isotropy of the stiffness of the graphene sheet, all results obtained for nanotubes with
the chirality index $(m, 0)$ will be valid for all other types of nanotubes with the same diameter $D$.
}

\section{Steady states of two interaction nanotubes}
Nanotubes exhibit a high longitudinal (axial) and relatively low transverse (radial) stiffness.
Because of this, a nanotube with a quite large diameter can undergo the transition from a hollow
cylindrical shape to a collapsed state
\cite{chopra1995,gao1998,xiao2007,baimova2015}
due to the nonvalent interaction of its layers.
Let us consider the possible stationary states of a system of two interacting nanotubes.

Let the node coordinates of the $k$-th nanotube ($k$-th cyclic chain) with chirality
index  $(m,0)$ be defined by $2N$-dimensional vector ${\bf x}_k=\{ {\bf u}_{k,n}\}_{n=1}^N$,
where $N=2m$ and $k=1,~2$. Then, the energy of nanotube deformation is
\begin{equation}
P_1({\bf x}_k)=\sum_{n=1}^N[V(R_n)+U(\theta_n)+W_0(y_n)+\frac12\sum^N_{l=1\above 0pt |l-n|>4} W_1(r_{n,l})].
\label{f6}
\end{equation}
The potential energy of nanotube system is
\begin{equation}
E=P_1({\bf x}_1)+P_1({\bf x}_2)+P_2({\bf x}_1,{\bf x}_2),
\label{f7}
\end{equation}
where the function
$$
P_2({\bf x}_1,{\bf x}_2)=\sum_{n_1=1}^N\sum_{n_2=1}^N W_1(r_{1,n_1;2,n_2})
$$
determines the energy of interaction between cyclic chains,
$
r_{1,n_1;2,n_2}=|{\bf u}_{1,n_1}-{\bf u}_{2,n_2}|
=[(x_{2,n_2}-x_{1,n_1})^2+(y_{2,n_2}-y_{1,n_1})^2]^{1/2}
$
is the distance between chain nodes.
\begin{figure}[tb]
\begin{center}
\includegraphics[angle=0, width=1.0\linewidth]{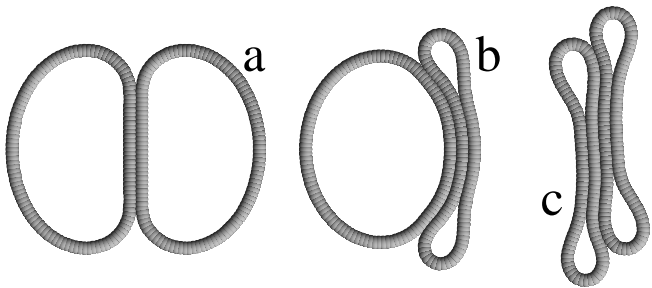}
\end{center}
\caption{\label{fig02}\protect
The stationary states of two interacting isolated nanotubes (60,0) (the number of chain nodes $N=120$):
(a) open, (b) open and collapsed, (c) collapsed nanotubes.
State energies $E=-1.021,~-1.223,~-0.740$~eV,
nanotube cross-sectional areas $S=30.84$, 20.52, 7.72~nm$^2$.
}
\end{figure}
\begin{figure}[tb]
\begin{center}
\includegraphics[angle=0, width=1.0\linewidth]{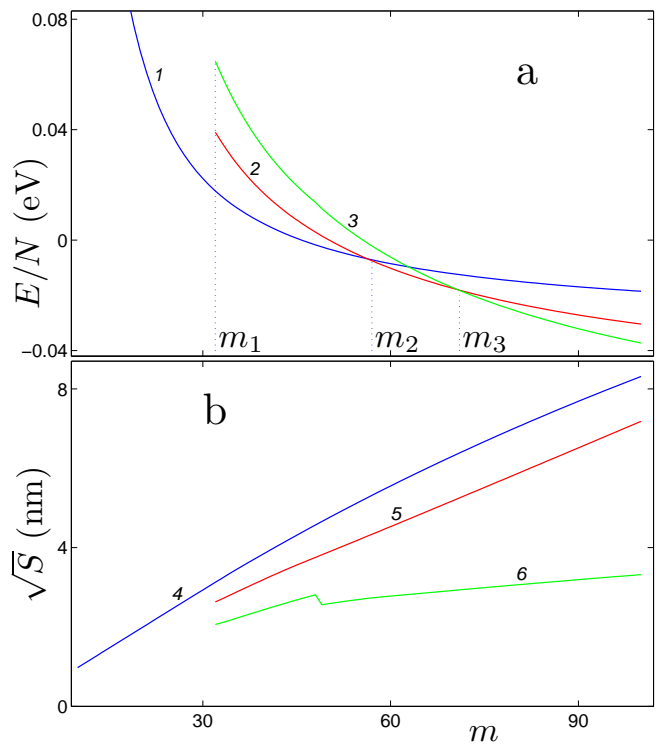}
\end{center}
\caption{\label{fig03}\protect
The dependence of (a) the energy $E/N$ and of (b) the square root of the cross-sectional area $\sqrt{S}$
of two interacting isolated nanotubes on the value of their chirality index $(m,0)$
(curves 1, 4  for open; 2, 5 for open and collapsed, and 3, 6 for collapsed nanotubes, $N=2m$).
The characteristic values of the index are $m_1=32$, $m_2=57$, $m_3=71$.
}
\end{figure}

To find the steady state of the CNT system, we must solve the energy minimization problem:
\begin{equation}
E\rightarrow\min: \{ {\bf x}_k\}_{k=1}^2.
\label{f8}
\end{equation}
Minimization problem (\ref{f8}) was solved numerically by a conjugated gradient method.
The stationary state of system of 2 CNTs $\{ {\bf x}_k^0\}_{k=1}^2$ is characterized
by the energy $E$ and by the cross-sectional area of the nanotubes $S$.

A characteristic picture of the three possible stationary states
{
 is shown in Fig.~\ref{fig02}.
Two CNTs with a chirality index (60,0) (the number of atoms in each chain $N=120$)
have three stationary stable states: (a) the state in which each nanotube is in the open state,
energy $E=-1.021$, cross-sectional area $S=30.84$;
(b) one nanotube is in the open and the other in the collapsed state, $E=-1.223$, $S=20.52$;
(c) all nanotubes are in a collapsed state $E=-0.740$~eV, $S=7.72$~nm$^2$.

The dependence of the energy $E$ and the cross-sectional area $S$ of two interacting nanotubes
on the value of their chirality index $(m,0)$ is shown in Fig.~\ref{fig03}.
The solution of the minimum problem (\ref{f8}) has demonstrated that at $m<32$ there is only one
stationary state in which all nanotubes are in the open state.
At $m\ge 32$, the system of two nanotubes already has three stationary states:
the state of open CNTs, the state of open and collapsed CNTs,
and the state of collapsed CNTs -- see Fig.~\ref{fig02}.
All states are stable. At the index $m<57$, the main state, i.e. the state with minimal energy,
is the state of open CNTs, at $57\le m<71$ -- the state of open and collapsed CNTs,
at $m\ge 71$ -- the state of collapsed CNTs.
Therefore, it should be expected that nanotube systems with a chirality index of $(m,0)$
at $m>32$ (with diameter $D>2.5$~nm) will have many stationary states,
which will be characterized by the portion of collapsed nanotubes.
\begin{figure}[tb]
\begin{center}
\includegraphics[angle=0, width=1.0\linewidth]{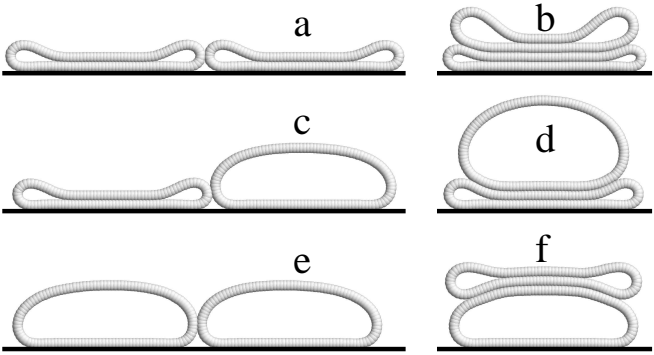}
\end{center}
\caption{\label{fig04}\protect 
The stationary states of two interacting nanotubes (60,0) (the number of chain nodes $N=120$)
located on a flat substrate:
(a) collapsed (longitudinal), 
(b) collapsed (vertical),
(c) collapsed and open (longitudinal),
(d) collapsed and open (vertical),
(e) open (longitudinal),
(f) open and collapsed (vertical arrangement).
State energies $E=-8.634$, -5.970, -8.009, -6.384, -7.409, -4.987~eV,
nanotube cross-sectional areas $S=6.36$, 7.15, 14.52, 18.03, 22.90, 14.31~nm$^2$.
Horizontal lines show the plane of the substrate.
}
\end{figure}
\begin{figure}[tb]
\begin{center}
\includegraphics[angle=0, width=1.0\linewidth]{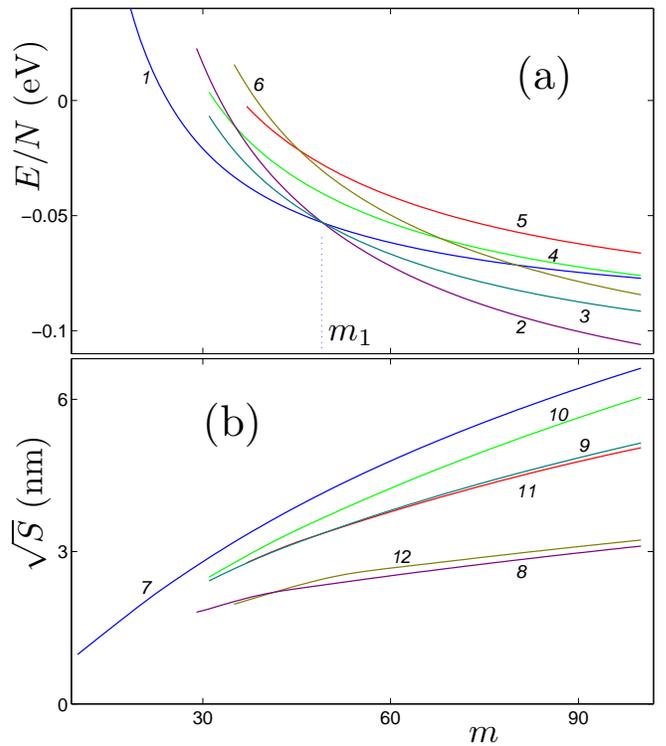}
\end{center}
\caption{\label{fig05}\protect 
The dependence of (a) the energy $E/N$ and of (b) the square root of the cross-sectional area $\sqrt{S}$
of two interacting nanotubes located on a flat substrate on the value of their chirality index $(m,0)$
(curves 1, 7  for open (longitudinal); 2, 8 for collapsed (longitudinal);
3, 9 for collapsed and open (longitudinal); 4, 10 for collapsed and open (vertical);
5, 11 for open and collapsed (vertical), and 6, 12 for  collapsed nanotubes(vertical arrangement), $N=2m$).
The characteristic value of the index is $m_1=49$.
}
\end{figure}

{
If interacting nanotubes are on a flat substrate, then six stationary stable states of 
two nanotubes are already possible -- see Fig.~\ref{fig04}.
They differ from each other not only in the open or closed (collapsed) state of the nanotubes 
but also in their mutual position relative to the substrate plane.
Thus, each nanotube can lie on the substrate, forming longitudinal arrangements (a), (c) and (e), 
or lie on top of each other, forming vertical arrangements (b), (d) and (f) in which only 
one of them is in contact with the substrate.

The dependence of the energy $E$ and the cross-sectional area $S$ of two interacting nanotubes
located on a flat substrate on the value of their chirality index $(m,0)$ is shown in Fig~\ref{fig05}.
The solution of the minimum problem (\ref{f8}) has demonstrated that the longitudinal arrangement 
of nanotubes on the substrate plane is always more energetically advantageous. 
At $m<49$, the most advantageous state is the longitudinal arrangement of open nanotubes, 
at $m>49$ -- the longitudinal arrangement of closed nanotubes.
}
\section{
Multilayer nanotube packing
}
Let us analyze the conformational changes of multilayer packages of single-walled nanotubes
during their transverse compression.
We consider a system of parallel $N_{xy}=N_x\times N_y $ nanotubes $(m,0)$ located between two
flat substrates ($N_x$ is the number of nanotubes in one layer parallel to the substrate,
$N_y$ is the number of layers) -- see Fig.~\ref{fig06}(a). Along the $x$ axis,
we will use periodic boundary conditions with the period $a_x$.
\begin{figure}[tb]
\begin{center}
\includegraphics[angle=0, width=1.0\linewidth]{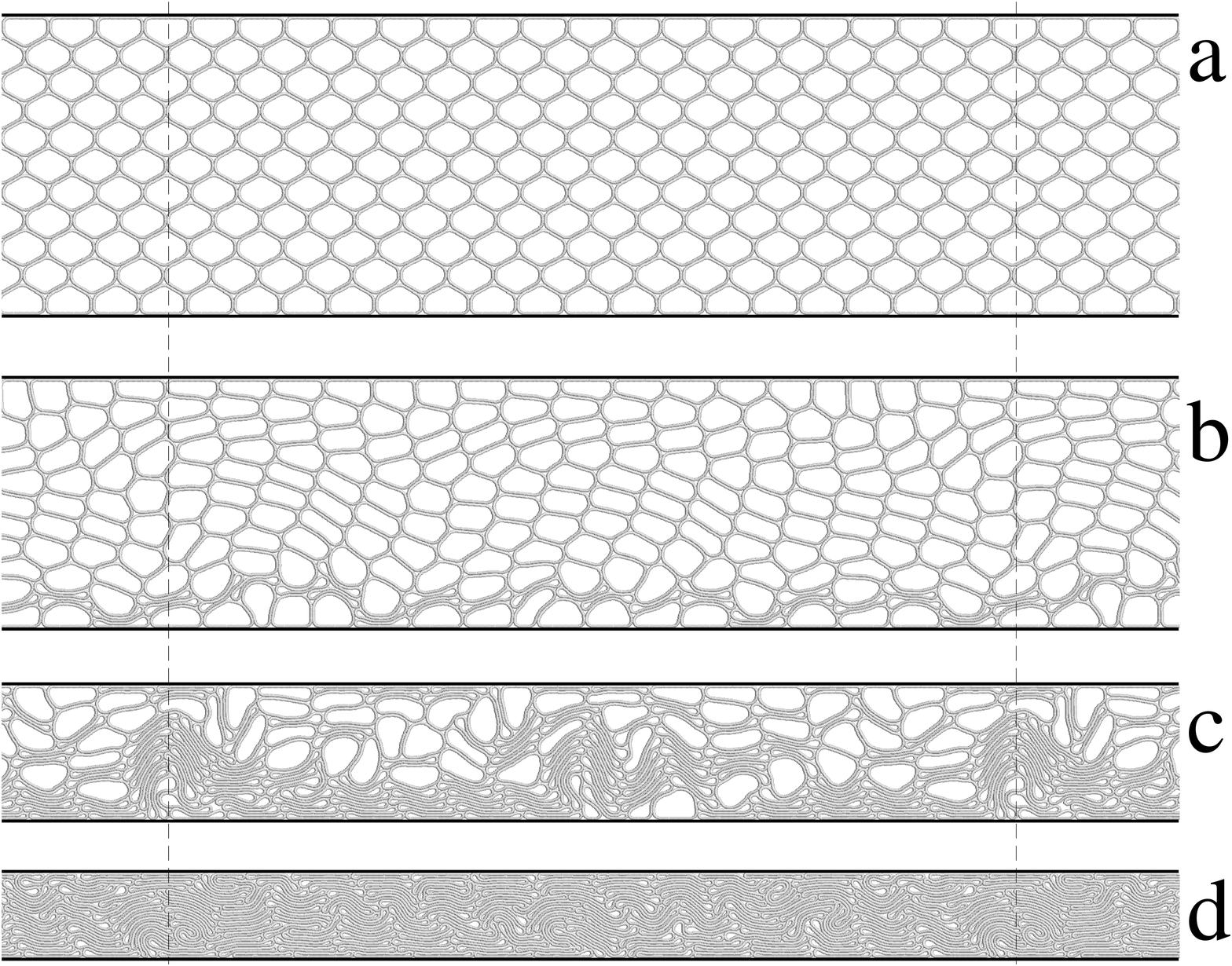}
\end{center}
\caption{\label{fig06}\protect
The picture of stationary states of a layered structure consisting
of $N_x\times N_y$ nanotubes $(60,0)$
(the number of nanotubes in one layer $N_x=18$, the number of layers $N_y=11$,
the number of atoms in each cyclic chain $N=120$) on a distance $h$ between the compressive
planes: (a) $h=36$ (state energy $E=-384.08$, pressure on the plane $P=0.00012$);
(b) $h=30$ ($E=-378.49$, $P=0.00042$); (c) $h=16$ ($E=-352.16$, $P=0.00084$);
(d) $h=10$~nm ($E=-229.35$~eV, $P=0.01156$~eV/\AA$^3$).
Horizontal lines show the compressive planes, while vertical lines show the boundaries
of the periodic cell (period $a_x=94.9$~nm).
}
\end{figure}
\begin{figure}[tb]
\begin{center}
\includegraphics[angle=0, width=1.0\linewidth]{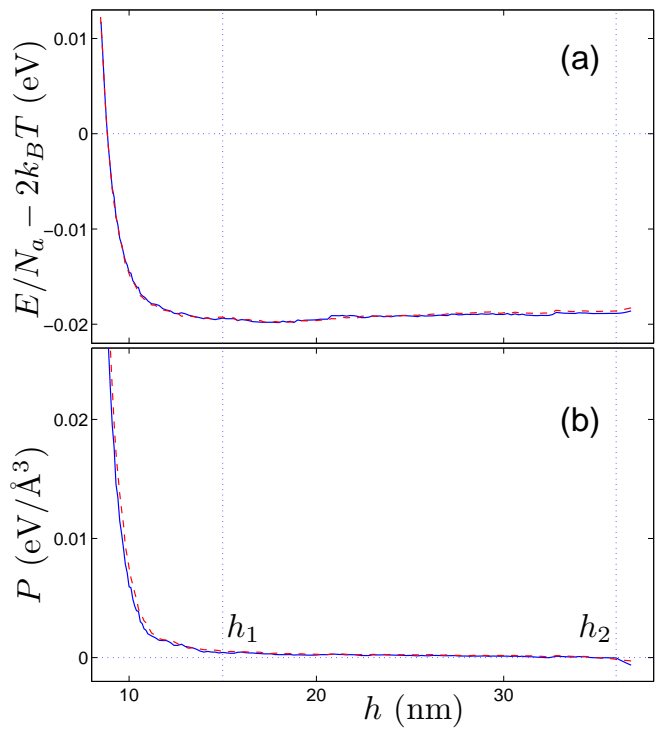}
\end{center}
\caption{\label{fig07}\protect
The dependence of (a) the energy $E$ normalized by the total number of atoms
of the system $N_a=N_xN_yN$ and (b) the pressure $P$ on the distance between
the compressive planes $h$. The characteristic values are $h_1=15$, $h_2=36$~nm.
The solid (blue) curves show the dependence for the states of the CNT system
at a temperature of $T=0$, the dotted (red) curves -- at $T=300$~K.
}
\end{figure}

The node coordinates  of the $k$-th nanotube ($k$-th cyclic chain) are
defined by $2N$-dimensional vector ${\bf x}_k=\{ {\bf u}_{k,n}\}_{n=1}^N$, $N=2m$,
$k=1,...,N_{xy}$. The energy of the nanotube deformation has the form
\begin{eqnarray}
P_1({\bf x}_k)=\sum_{n=1}^N[V(R_n)+U(\theta_n)+W_0(y_{k,n})\nonumber\\
+W_0(h-y_{k,n})+\frac12\sum^N_{l=1\above 0pt |l-n|>4} W_1(r_{n,l})],
\label{f9}
\end{eqnarray}
where $h$ is distance between the surfaces of flat substrates.
The potential energy of nanotube packing per unit cell (per period) is
\begin{eqnarray}
E=\sum_{k=1}^{N_{xy}}P_1({\bf x}_k)+\sum_{k_1=1}^{N_{xy}-1}\sum_{k_2=k_1+1}^{N_{xy}-1}
[P_2({\bf x}_{k_1},{\bf x}_{k_2})\nonumber\\
+P_2({\bf x}_{k_1},{\bf x}_{k_2}+a_x{\bf e}_x)],
\label{f10}
\end{eqnarray}
where vector ${\bf e}_x=\{(1,0)\}_{n=1}^N$.
\begin{figure}[tb]
\begin{center}
\includegraphics[angle=0, width=1.0\linewidth]{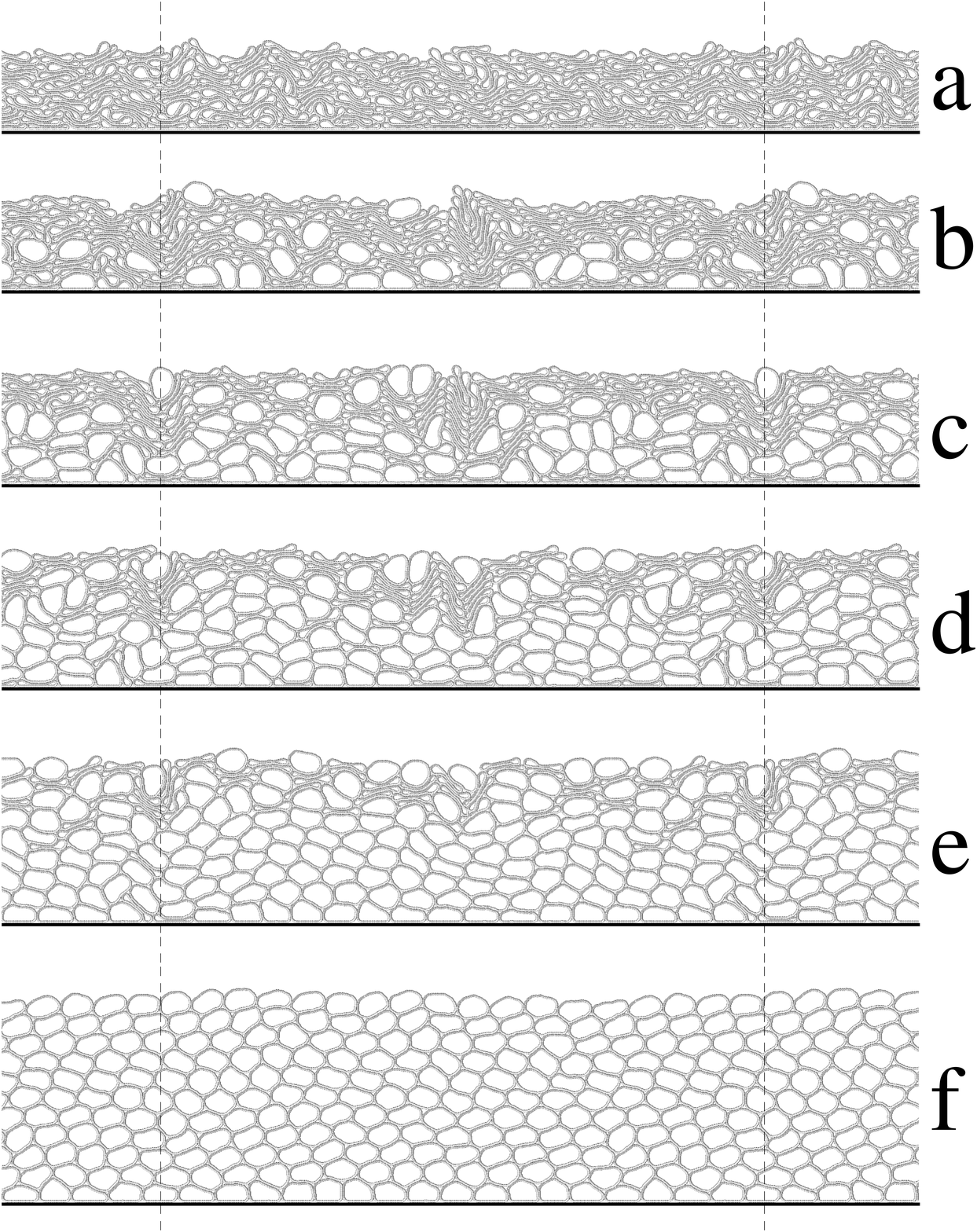}
\end{center}
\caption{\label{fig08}\protect
The picture of stable states of layered structures consisting of
$N_x\times N_y$ nanotubes $(60,0)$
(the number of nanotubes in one layer $N_x=18$, the number of layers $N_y=11$,
the number of atoms in each cyclic chain $N=120$) located on a flat substrate
with an average thickness of the structure (a) $h_y=12.98$, (b) 16.14, (c) 20.49,
(d) 24.90, (e) 30.14, (f) 35.10~nm.
The states are shown at a temperature of $T=300$~K.
Horizontal lines show the plane of the substrate, while vertical lines  --
the boundaries of the periodic cell (period $a_x=94.9$~nm).
}
\end{figure}
\begin{figure}[tb]
\begin{center}
\includegraphics[angle=0, width=1.0\linewidth]{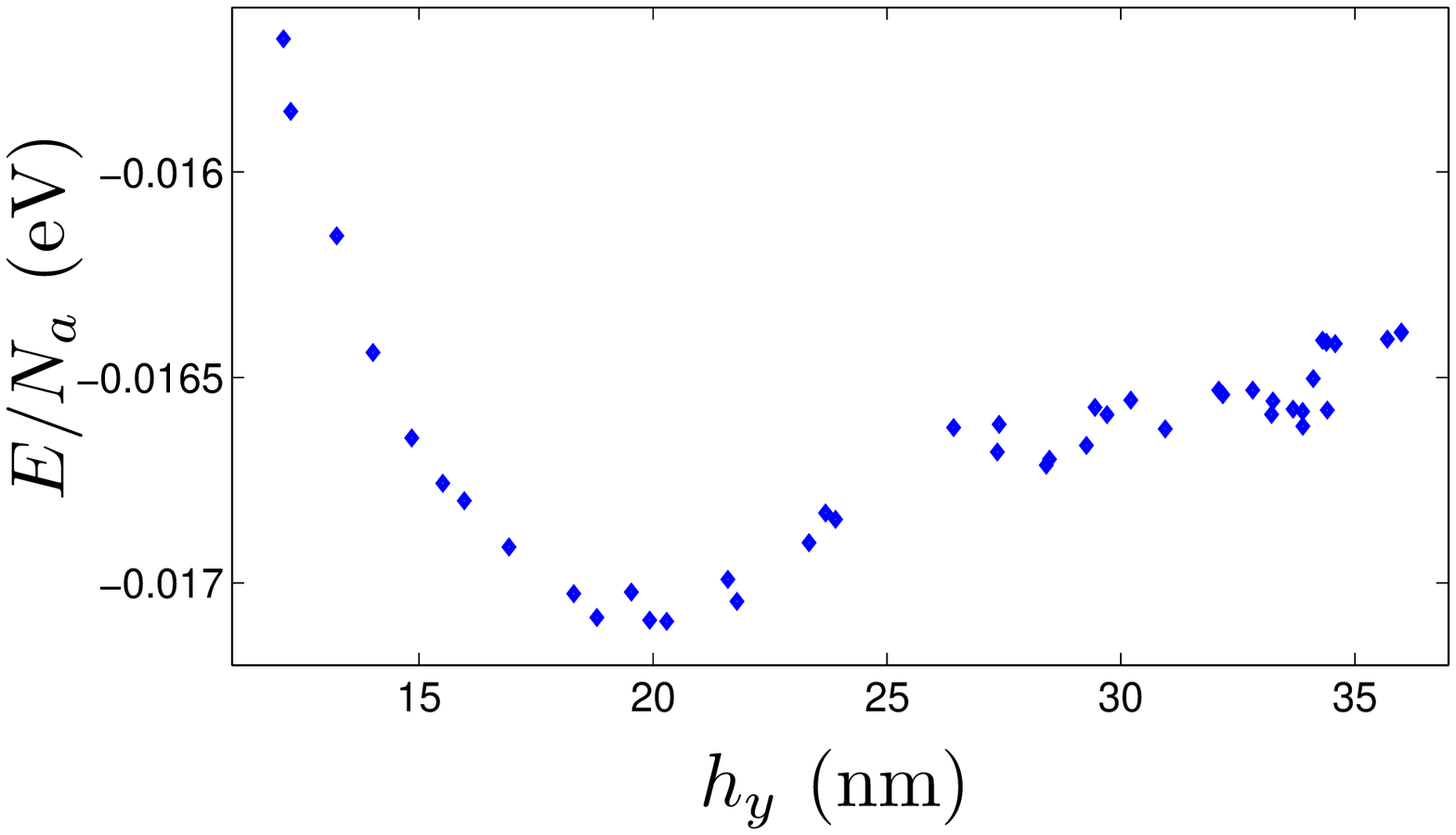}
\end{center}
\caption{\label{fig09}\protect
The dependence of the energy of stationary states of layered structures
consisting of $N_x\times N_y$
nanotubes (60,0) (the number of nanotubes in one layer $N_x=18$, the number of layers $N_y=11$,
the number of atoms in each cyclic chain $N=120$) located on a flat substrate
on the average thickness of the structure $h_y$.
}
\end{figure}

Let us take for certainty the index $m=60$ (the number of nodes in the chain $N=120$),
the number of nanotube layers $N_y=11$, the number of nanotubes in one layer $N_x=18$,
the unit cell size (period) $a_x=94.9$~nm.
The stationary state of a multilayer system of nanotubes was found numerically
as a solution to the minimum energy problem
\begin{equation}
E\rightarrow\min:\{ {\bf x}_k\}_{k=1}^{N_{xy}}
\label{f11}
\end{equation}
for fixed values of the distance between the layers $h$ and the unit cell size $a_x$.

Let us first take $h=36.8$~nm and get the stationary state of the multilayer packaging
of open nanotubes -- see Fig.~\ref{fig06}.
Then we will gradually reduce the distance between the layers $h$.
Compression will lead to the collapse of a part of the nanotubes.
The number of collapsed nanotubes increases monotonically with a decrease in the distance
between the layers. At $h=10.9$~nm, almost all nanotubes will already be in the collapsed state.

Each stationary state $\{ {\bf x}_k^0\}_{k=1}^{N_{xy}}$ is characterized by the energy $E$
and by the pressure on the substrate plane
$$
P=\frac{1}{3a_xr_c}\sum_{k=1}^{N_{xy}}[W_0'(h-y_{k,n}^0)-W_0'(y_{k,n}^0)].
$$
The dependence of $E$ and $P$ on $h$ is shown in Fig.~\ref{fig07}.
As can be seen from the figure, the range of values of $h$ can be divided into three zones.
At $h>h_2=36$~nm, the walls stretch a multilayer system of open nanotubes.
The compression of the nanotube system begins to occur only when the distance between the layers $h<h_2$.
In the range of values $h_1<h<h_2$, $h_1=15$, compression leads only to a decrease in the energy
of the system $E$ and to a weak linear increase in pressure $P$. The energy and the pressure begin
to increase sharply at $h<h_1$. The minimum energy is achieved when the compression $h=17.4$~nm.

In order to obtain the free state of the multilayer packaging of nanotubes,
we will remove one substrate and solve the minimum problem (\ref{f11}) using
the compressed stationary state as the starting point in the conjugate gradient method.
As a result, we get a stationary state of the multilayer packaging of nanotubes
$\{ {\bf x}_k^0\}_{k=1}^{N_{xy}}$ at zero pressure.
Each such state will be characterized by the energy $E$ and by the average thickness of the packaging
$$
h_y =\frac{2}{N_a}\sum_{k=1}^{N_{xy}}\sum_{n=1}^N y_{k,n}^0,
$$
where the total number of atoms of CNT system $N_a=N_{xy}N=N_xN_yN$.

A characteristic view of stationary uncompressed multilayer nanotube packages
is shown in Fig. \ref{fig08}.
Calculations have shown that the multilayer system of nanotubes (60,0) is a multistable,
its stationary states can have an average thickness of $12<h_y<36$~nm -- see Fig.~\ref{fig09}.
Thus, the thickness of different packages may differ by three times.
All stationary packages are stable to thermal fluctuations and differ in the ratio of the number
of open to the number of collapsed nanotubes (the more is the number of open nanotubes,
the greater is the package thickness).
The thickest is the package with all open nanotubes (see Fig.~\ref{fig08}f), its thickness is $h_y=35.10$~nm,
the thinnest is the package with all collapsed nanotubes (see Fig.~\ref{fig08}a), its thickness is $h_y=12.98$~nm.

The dependence of the energy of $N_y$-layer packages on their thickness is shown in Fig.~\ref{fig09}.
As can be seen from the figure, the energy of the packages weakly depends on their thickness.
The minimum energy $E/N_a=-0.017$~eV is achieved at thickness $h_y=20$~nm.

{
Note that all the stationary states of nanotube arrays obtained using the 2D model will correspond 
to the stationary states in the 3D model if the lengths of the nanotubes significantly 
exceed their diameters (if $L\gg D$).
}

{
\section{Accounting of thermal vibrations}

To check the stability of the stationary states of multilayer nanotube packages, molecular dynamic 
modeling was performed at a temperature of $T=300$ K. 
A system of Langevin equations was numerically integrated to simulate thermal oscillations
\begin{equation}
M\ddot{\bf x}_k=-\frac{\partial E}{\partial {\bf x}_k}-\Gamma M\dot{\bf x}_k-\Xi_k,~~k=1,...,N_{xy},
\label{f12}
\end{equation} 
where ${\bf x}_k$ is $2N$-dimensional vector giving the coordinates of the $k$th nanotube, 
$E$ is potential energy of molecular system (\ref{f10}), $M$ is the  mass of carbon atom,
$Gamma=1/t_r$ is the friction coefficient (the relaxation time $t_r=1$~ps),
$\Xi_k=\{(\xi_{k,n,1},\xi_{k,n,2})\}_{n=1}^N$ is $2N$-dimensional vector of normally
distributed random Langevin forces with the following correlations:
$$
\langle\xi_{k_1,n_1,i}(t_1)\xi_{k_2,n_2,j}(t_2)\rangle=2Mk_BT\Gamma\delta_{k_1k_2}\delta_{n_1n_2}\delta_{ij}\delta((t_2-t_1)
$$
($k_B$ is Boltzmann constant, $T$ is temperature of the Langevin thermostat).

As an initial condition for the equations of motion (\ref{f12}) we take the stationary state 
of nanotube packaging. 
Numerical integration of the system of equations of motion (\ref{f12}) showed that
all stationary states of the nanotube system are stable to thermal fluctuations.
As can be seen from Fig. \ref{fig07}, thermal fluctuations not lead 
to a significant change in the type of dependencies $E(h)$ and $P(h)$.

\section{Numerical methods}
The minimum energy problems (\ref{f8}), (\ref{f11}) were solved numerically 
by the conjugate gradient method \cite{fletcher1964}.
The system of equations of motion (\ref{f12}) was integrated numerically using 
the velocity form of the Verlet difference scheme \cite{verlet1967}.
All the simulation programs were written in FORTRAN.
Programs RasMol and RasTop were used for the visualization of obtained results. 
The main calculations were carried out on supercomputers of the Joint Supercomputer Center, 
Russian Academy of Sciences.
}

\section{conclusions}

The simulation has shown that the multilayer packaging of identical single-walled nanotubes
with a diameter $D>2.5$~nm located on a flat substrate is a multistable system.
The system has many stationary states, which are characterized by the portion of collapsed nanotubes.
The thickness of the package monotonically decreases with an increase in the portion of such nanotubes.
For nanotubes with a chirality index (60,0), the thickness of the 11-layer package can vary
from 12 to 36~nm, depending on the portion of collapsed nanotubes.
All stationary states of the package are stable to thermal fluctuations at $T=300$~K.
The transverse compression of the package is not elastic, it only, due to the collapse
of a part of the nanotubes, transfers the package from one stationary state to another
with a smaller width.

{
Compression of CNT packages will occur elastically only for $(m,0)$ nanotubes with 
the index $m\le 32$ (for CNT with diameter $D\le 2.5$~nm). If we take any strongly 
compressed stationary state of a multilayer package and allow free movement of 
the compressing walls, the package will expand and return to its basic uncompressed state. 
Such multi-layer packages will give the best protection against vibrations.
}

\medskip
\textbf{Acknowledgements} \par 
The work was supported by the Russian Foundation for Basic Research
and by the Department of Science
and Technology, Ministry of Science and Technology, Government
of India, within scientific project no. 19-58-45036.
Computational resources were provided by the Joint Supercomputer
Center, Russian Academy of Sciences.


\end{document}